\documentclass[usenatbib,preprint]{emulateapj}

\usepackage{graphicx}
\usepackage {natbib,aas_macros}
\usepackage {bm}
\usepackage {amsmath,amssymb}
\usepackage {setspace}
 
\newcommand{\cm}{~{\rm g~cm}^{-3} }

\begin{document}
\title{Bimodality of circumstellar disk evolution induced by Hall current}

\author{
Y. Tsukamoto\altaffilmark{1,2},  K. Iwasaki\altaffilmark{2,3}, S. Okuzumi\altaffilmark{4}, M. N. Machida\altaffilmark{5}, and  S. Inutsuka\altaffilmark{2}
}
\altaffiltext{1}{
Laboratory of Computational Astrophysics, RIKEN, Saitama, Japan }
\altaffiltext{2}{
Department of Physics, Nagoya University, Aichi, Japan  }
\altaffiltext{3}{
Department of Environmental Systems Science,
Faculty of Science and Engineering, Doshisha University, Kyoto, Japan }
\altaffiltext{4}{Department of Earth and Planetary Sciences, Tokyo Institute of Technology, Tokyo, Japan }
\altaffiltext{5}{Department of Earth and Planetary Sciences, Kyushu University, Fukuoka, Japan }

\begin{abstract}
The formation process of circumstellar disks 
is still controversial because of the interplay of complex physical processes 
that occurs during the gravitational collapse of prestellar cores.
In this study, we investigate the effect of 
the Hall current term on the formation of the circumstellar disk
using three-dimensional simulations.
In our simulations, all non-ideal effects 
as well as the radiation transfer are considered.
The size of the disk is significantly affected by 
a simple difference in the inherent properties of the prestellar core, namely
whether the rotation vector and the magnetic field are 
parallel or anti-parallel.
In the former case, only a very small 
disk ($< 1$ AU) is formed. On the other hand, in the latter case, 
a massive and large ($>20$ AU) disk is formed in the early phase of 
protostar formation.
Since the parallel and anti-parallel properties 
do not readily change, we expect that 
the parallel and anti-parallel properties
are also important in the subsequent disk evolution 
and the difference between the two cases is maintained or enhanced. 
This result suggests that the disk size distribution
of the Class 0 young stellar objects is bimodal.
Thus, the disk evolution can be categorized 
into two cases and we may call 
the parallel and anti-parallel systems as
{\it Ortho-disk} and {\it Para-disk}, respectively.
We also show that the anti-rotating envelopes 
against the disk-rotation appear with a size of $\gtrsim 200$ AU.
We predict that the anti-rotating envelope 
will be found in the future observations.
\end{abstract}


\section{Introduction}
\label{intro}


Circumstellar disks are born around protostars 
in the course of the self-gravitational 
collapse of the molecular cloud cores. 
The angular momentum evolution during the collapse is critical
for the disk formation 
because the disks are supported by centrifugal force.

The magnetic field plays a central 
role for the angular momentum evolution.
During the gravitational collapse, a toroidal magnetic field is 
created by the rotation, and 
the magnetic tension decelerates the gas rotation, removing
the angular momentum.
This effect is known as magnetic braking \citep{1979ApJ...230..204M}. 
Previous studies using ideal magnetohydrodynamics (MHD) simulations
have shown that,
for a typical magnetic field strength, the disk formation is 
completely suppressed in the 
early Class 0 phase 
\citep{2008ApJ...681.1356M,2008A&A...477....9H, 2014MNRAS.437...77B}.

Meanwhile, the observations of Class 0 
young stellar objects (YSOs) 
showed that a relatively large circumstellar disk
($r \sim 50 $ AU) exists around young protostars
\citep{2014ApJ...796..131O,2014Natur.507...78S,2015arXiv150305189T}.
These observations indicate that some very young protostars have
relatively large circumstellar disks with a size of $r>10$ AU, suggesting
a disagreement between the previous theoretical works and 
the observations.

Possible physical mechanisms for 
resolving the discrepancy between theory and observation are
the Ohmic and ambipolar diffusion
\citep{2009ApJ...698..922M,
2014MNRAS.438.2278M,2015ApJ...801..117T,2015arXiv150304901T}.
Previous works following the formation of protostars showed that 
a small disk with size of $r\lesssim 1$ AU is formed around the protostar
when the magnetic diffusion are considered
\citep{2013ApJ...763....6T,2015arXiv150304901T}.
However, the formation of a disk with a size of $r\gtrsim10$ AU 
at the early phase of protostar 
formation in a typically magnetized cloud core 
is still highly difficult even with these effects.

The effect of the Hall current term 
is the least studied effect in the context of disk formation.
The Hall current term generates a toroidal magnetic field 
from a poloidal magnetic field and directly
affects the magnetic tension that
determines the magnetic braking efficiency.
The Hall current term in MHD equations 
is not invariant against the global inversion of the magnetic field 
\citep{1999MNRAS.303..239W,2011ApJ...733...54K,
2011ApJ...738..180L,2012MNRAS.422..261B} 
and its effect changes
depending on whether rotation vector and magnetic 
field of host cloud core are parallel or anti-parallel.
When bf the Hall diffusion coefficient is negative (this is true for
$\rho \lesssim 10^{-11} \cm $ in our model, see figure \ref{eta})
and the rotation vector and the magnetic 
field are anti-parallel, the Hall current term weakens 
the magnetic braking.
Meanwhile, the Hall current term strengthens 
the magnetic braking in the parallel case.
Despite the possible importance of the Hall current term in the disk 
evolution,
it is still unclear how the Hall current term affects the formation 
of the circumstellar disk because the previous numerical studies 
 \citep{2011ApJ...733...54K,2011ApJ...738..180L} neglected
the first core evolution phase, which plays an important role in
the disk formation
\citep{2011MNRAS.413.2767M,2012A&A...541A..35D,2015ApJ...801..117T,2015arXiv150304901T}
and simplified the radiative transfer. 
Three-dimensional simulations are also
necessary for investigating non-axisymmetric effects.

In this paper,
we performed three-dimensional simulations 
starting from prestellar cloud cores. 
Our numerical simulations include
all non-ideal MHD effects as well as the radiative transfer.
The simulations were conducted until
the birth of the protostar.
We did not use any sink technique 
for the center and hence our simulations do not 
suffer from numerical artifacts introduced by 
sink particle or the inner boundary 
which may artificially change the 
formation and evolution of the disk 
\citep{2014MNRAS.438.2278M}.



\begin{figure*}
\includegraphics[width=70mm]{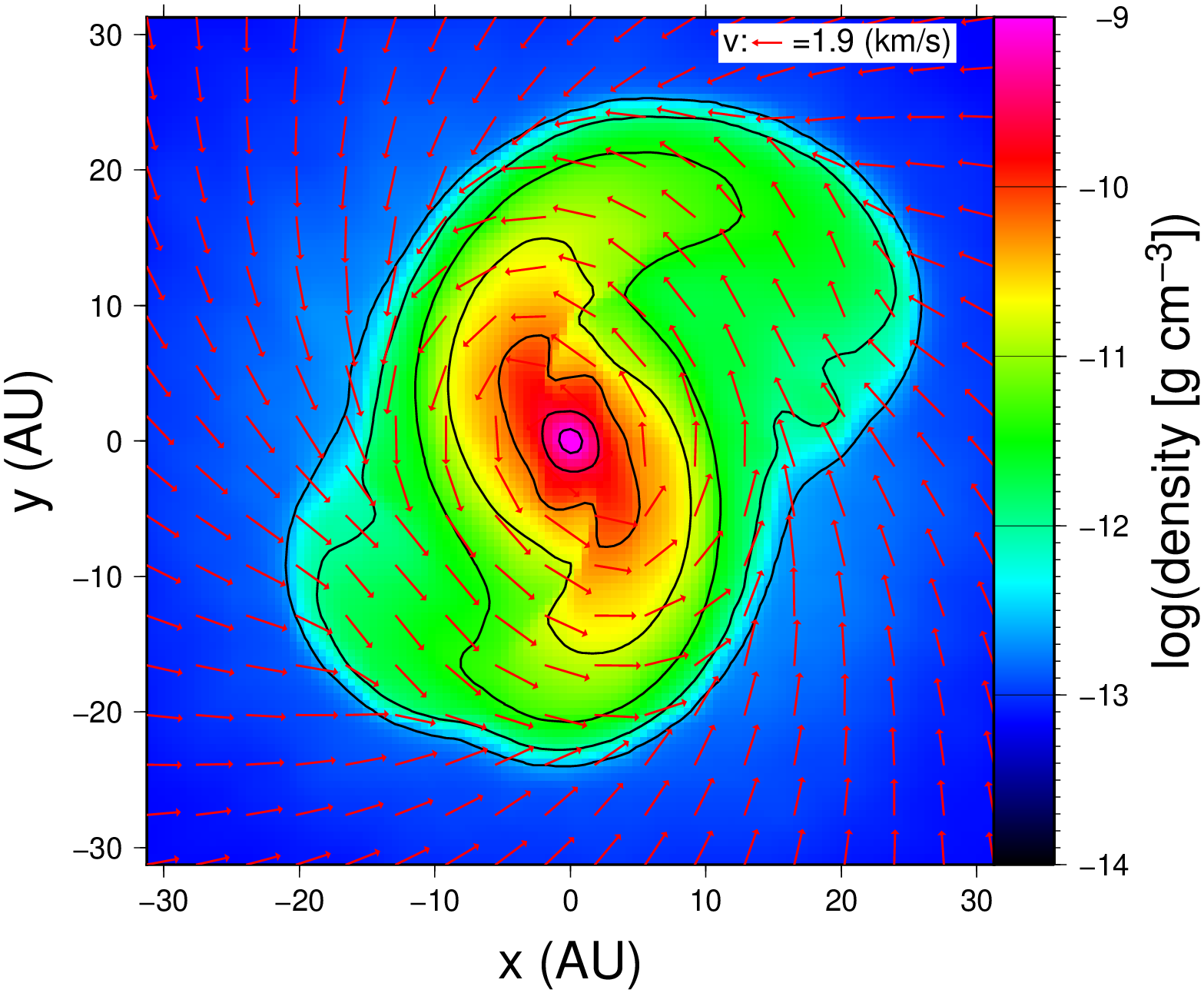}
\includegraphics[width=70mm]{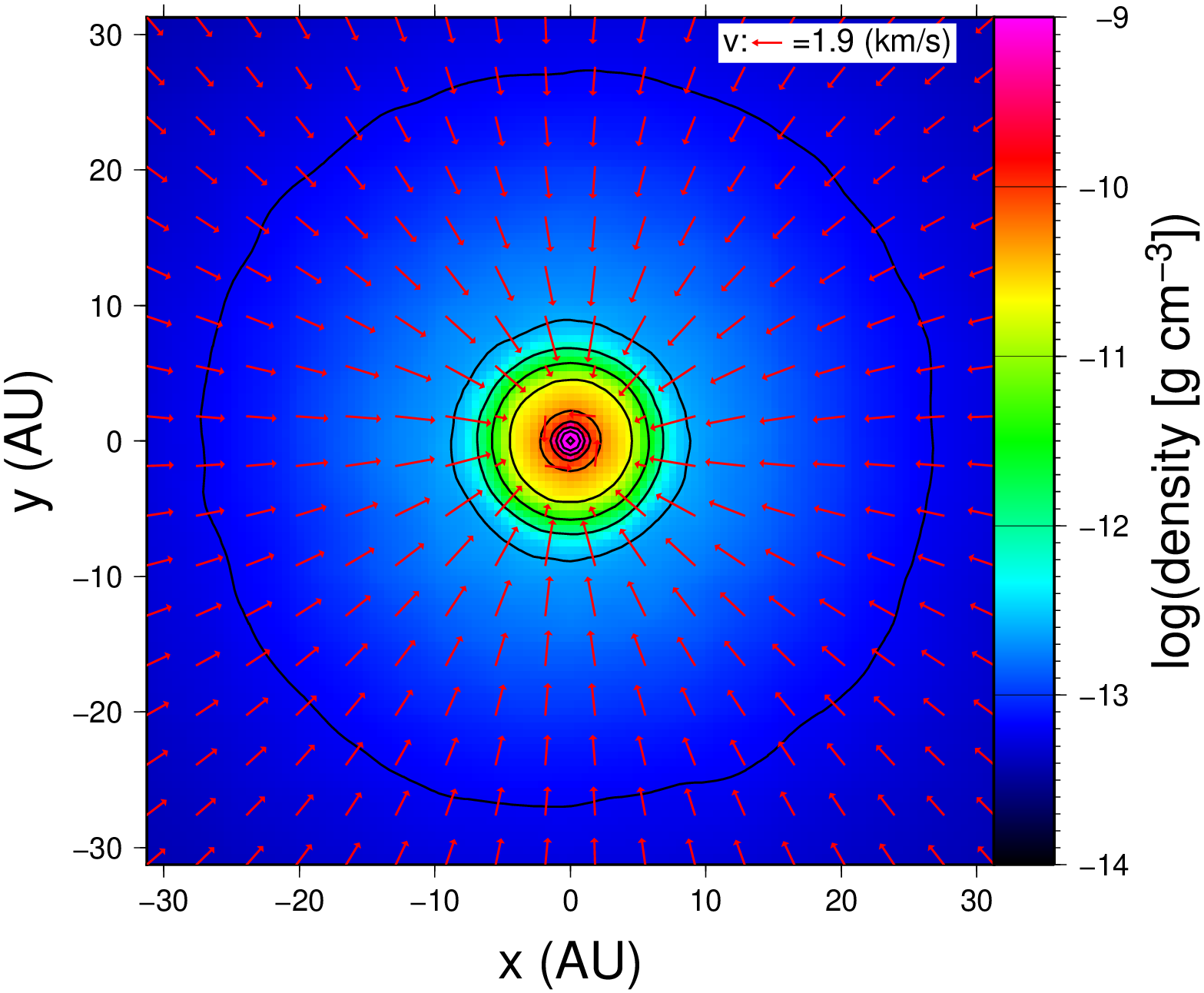}

\includegraphics[width=70mm]{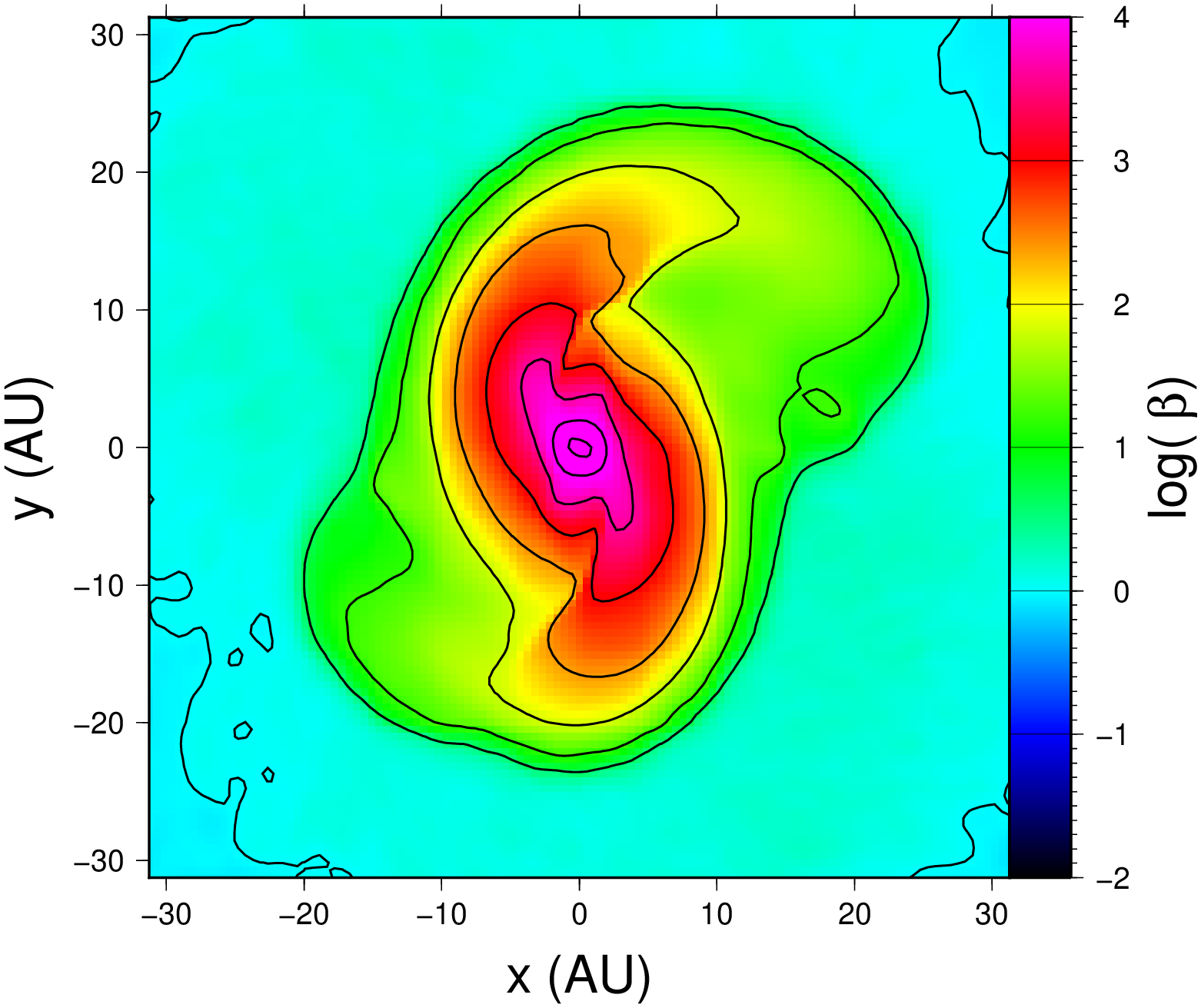}
\includegraphics[width=70mm]{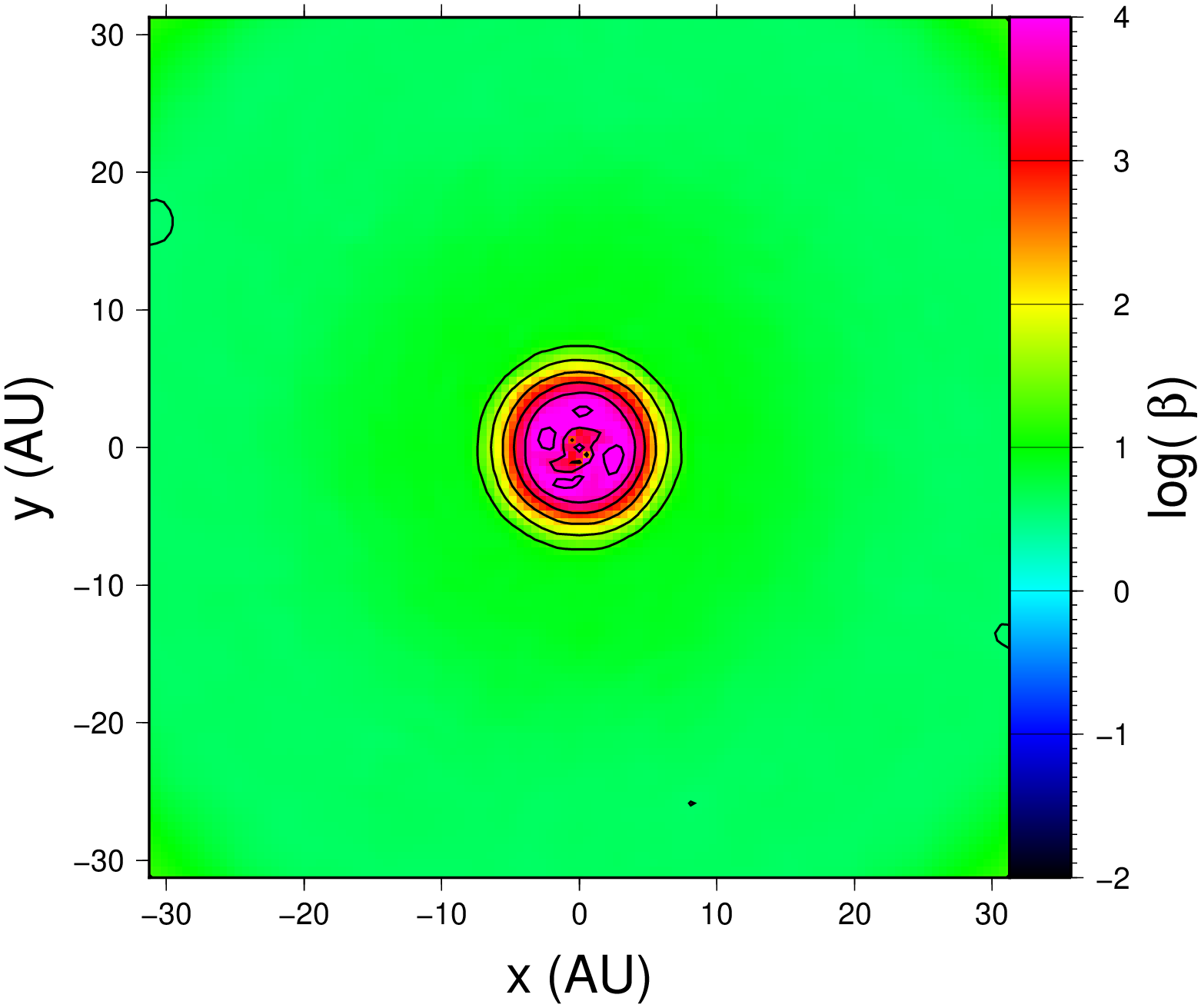}

\caption{
Cross sections of the density (top panels) 
and the plasma $\beta$ (bottom panels) in the x-y plane.
The left panels show the results of Model Para, while
the right panels show those of Model Ortho.
The central densities at these snapshots are $10^{-9} \cm$ 
and $10^{-2} \cm$ for the left and right panels, respectively.
}
\label{xy_map}
\end{figure*}

\section{Numerical Method and Initial Conditions}
\label{method}
In this study, 
we solved the non-ideal radiation magneto-hydrodynamics 
equations with self-gravity.
The numerical method, except for the Hall current term, is the same as that 
used in our previous study \citep{2015arXiv150304901T}.
The ideal MHD part was solved using the methods proposed by
\citet{2011MNRAS.418.1668I,2013ASPC..474..239I}.
The radiative transfer was treated with the methods of
\citep{2004MNRAS.353.1078W,2005MNRAS.364.1367W}.
We treated the Ohmic and ambipolar diffusion with the method described by
\citet{2013MNRAS.434.2593T} and
\citet{2014MNRAS.444.1104W}, respectively.
Both diffusion processes were accelerated by super time stepping 
\citep{Alexiades96}.

For this study, we newly implemented the 
Hall current term
according to \citet{2014MNRAS.444.1104W}.
The Hall current term couples with the ideal terms and 
changes the phase velocity. Thus, it is unclear that
the sub-cycle method, which is often used 
for the diffusion terms \citep{2011PASJ...63..555M,2015arXiv150304901T},
is valid for the Hall current term.
Therefore, in our simulations, the ideal MHD term and the Hall current term
are updated with the same time-step.
The smaller time-step of either 
the ideal term or the Hall current term, 
$\Delta t_{\rm Hall}=C_{\rm hall}h^2/(4 \pi |\eta_{\rm H}|)$ \citep{2002ApJ...570..314S} is used for the update.
Here, $C_{\rm hall}=0.4$ is the Courant-Friedrichs-Levy  
number for the Hall term, and $h$ is the smoothing length.

We conducted the numerical tests for the Hall current term and 
confirmed that the scheme can correctly calculate the whistler mode in the 
linear wave propagation test.
Furthermore, we conducted gravitational collapse tests of non-rotating
cloud cores and confirmed that the rotation amplitude induced 
by the Hall current term
does not depend on the direction of the magnetic field.
We adopted the equation of state (EOS), and 
dust and gas opacity tables
from \citet{2013ApJ...763....6T}, \citet{2003A&A...410..611S}, and \citet{2005ApJ...623..585F}, respectively.
We employed the resistivity table used 
in our previous works \citep{2009ApJ...698.1122O,2015arXiv150304901T}
in which the fixed dust grain size of $a=0.035 \mu m$ and fixed 
cosmic-ray ionization rate of $\xi_{\rm CR}=10^{-17} s^{-1}$ are assumed.

We modeled the initial cloud core with an isothermal uniform gas sphere
using about $3 \times 10^6$ particles.
The mass and temperature of the initial core
are 1 $M_\odot$ and 10 K, respectively.
Initially, the core has a radius of $R = 3.0\times 10^3 $ AU and
is rigidly rotating
with an angular velocity of $\Omega_0=2.2 \times 10^{-13}~{\rm s^{-1}}$.
The initial magnetic field is uniform and parallel to the rotation (z-) axis 
with a magnitude of $B_0=1.7\times 10^2 {\rm \mu G}$. 
The corresponding initial mass-to-flux ratio 
relative to the critical value is $\mu=(M/\Phi)/(M/\Phi)_{\rm crit}=4$ where
$\Phi=\pi R^2 B_0$ and  $(M/\Phi)_{\rm crit}=(0.53/3 \pi)(5/G)^{1/2}$
\citep{1976ApJ...210..326M}.

We conducted three simulations, Model Para, Ortho, and NoHall. 
The magnetic field and the rotation vector are initially perfectly
parallel in Model Ortho, NoHall
and anti-parallel in Model Para. The model NoHall does not include the 
Hall current term. Other parameters are the same in the models.
The runtime of Models Para, Ortho, and NoHall
was about $3.7\times10^5, 1.4\times10^5, 1.6\times10^4$ CPU hours, 
respectively with XC30 in NAOJ.

A boundary condition is imposed at $R_{\rm out}=0.995 R$,
and the particles with $r>R_{\rm out}$ rotate with an initial
angular velocity. Thus, the gas is confined in a rigidly rotating shell.
This boundary is very similar to that used 
in \citet{2004ApJ...616..266M,2007ApJ...670.1198M} and also used in our 
previous work \citep{2015arXiv150304901T}.
In addition, a boundary condition for radiative
transfer is introduced by fixing 
the gas temperature to be 10 K when $\rho<2.0 \times 10^{-17} \cm$.

\section{Results}
In figure \ref{xy_map}, we show the structure at 
the center of the cloud core. 
The left and right panels show the result 
of Models Para and Ortho, respectively.
The central densities are $10^{-9} \cm$ 
for the left panels, which correspond to slightly before 
the protostar formation, and $10^{-2} \cm$ for the 
right panels, which is immediately after the protostar formation.

\begin{figure}
\includegraphics[width=60mm,angle=-90]{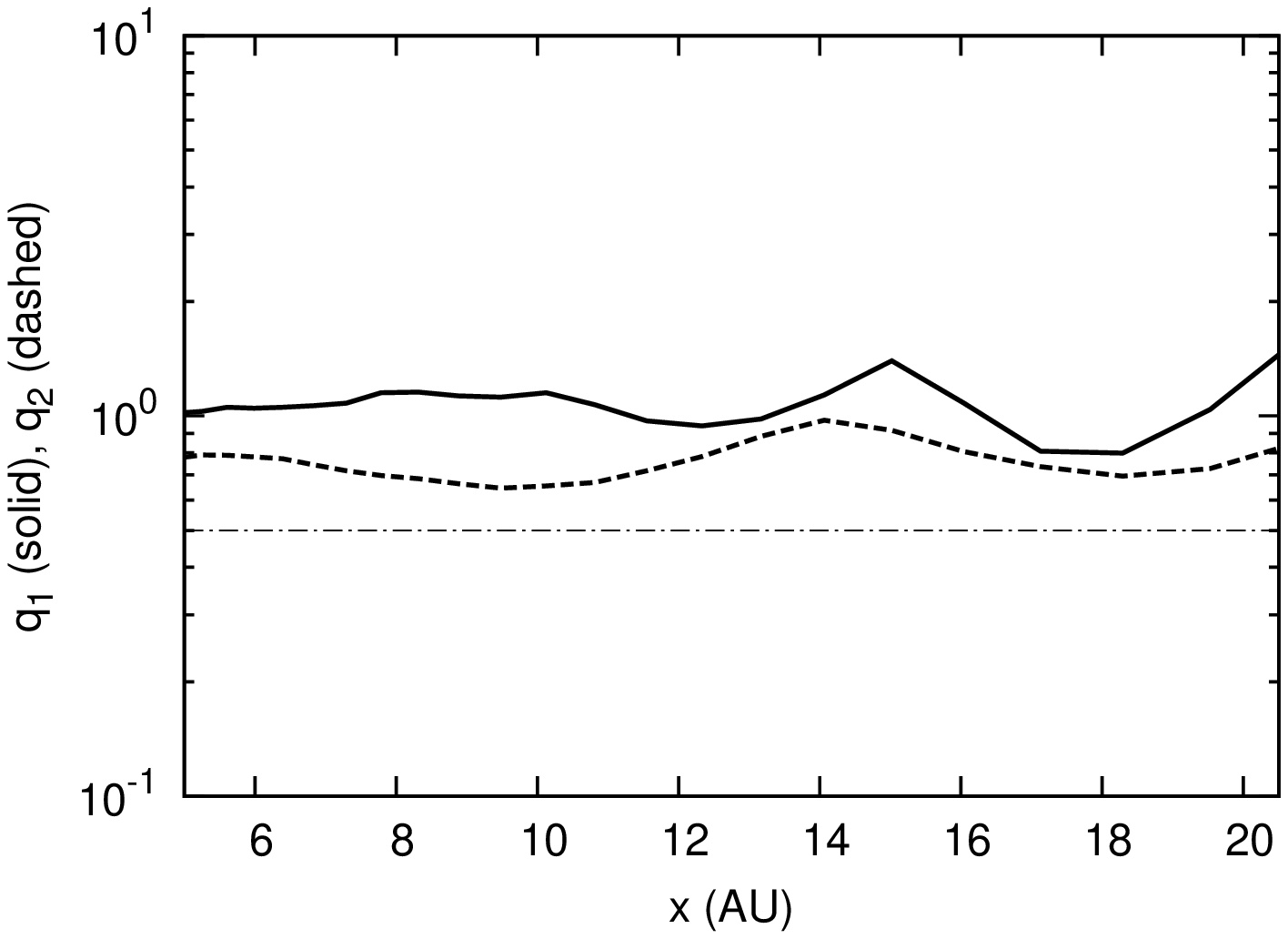}
\includegraphics[width=60mm,angle=-90]{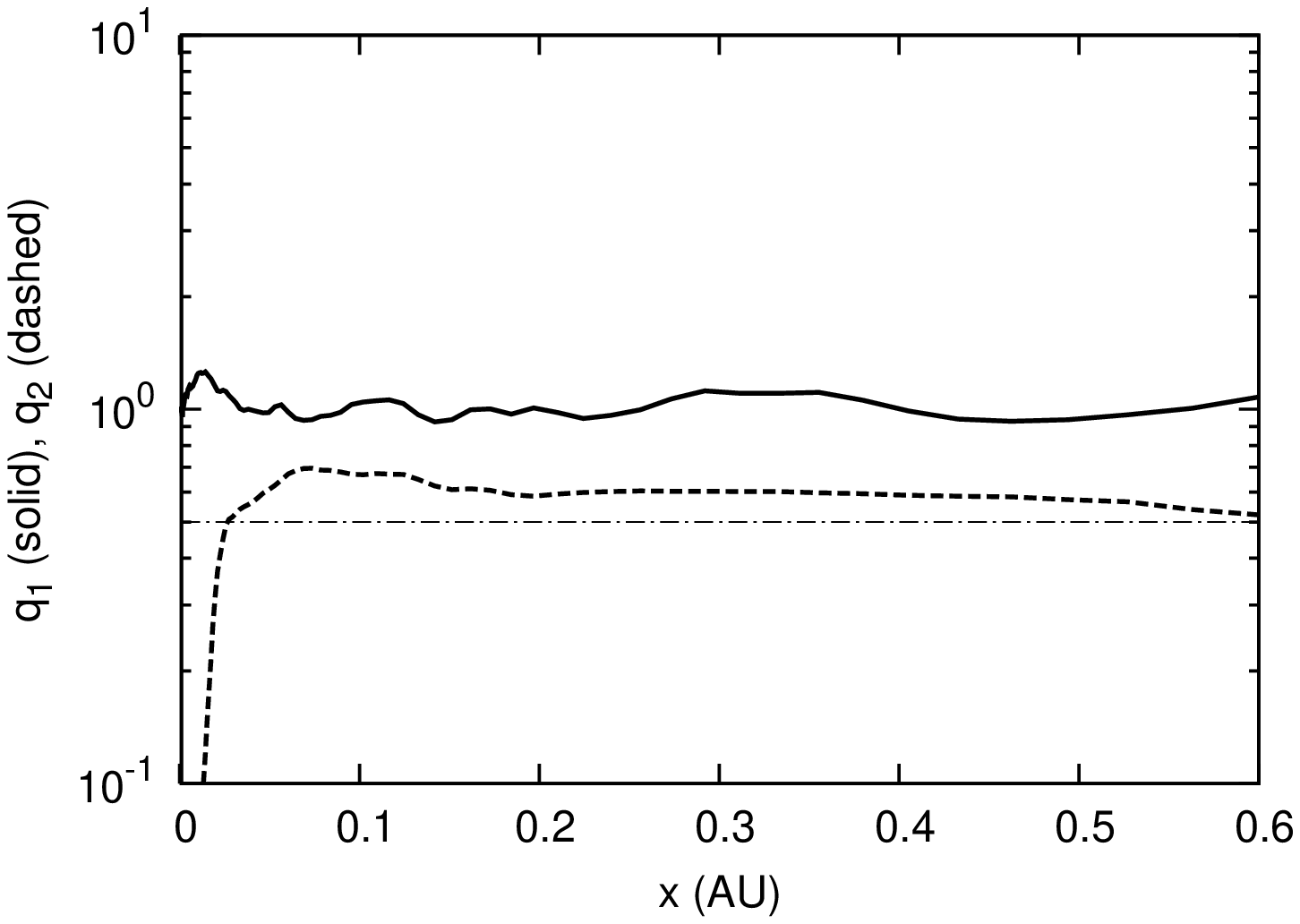}
\caption{
Solid lines show the ratio of the sum of the centrifugal 
force and the pressure gradient force
to the radial gravitational force,
$
q_1=|\frac{v_\phi^2/x+\nabla_r p/\rho}{\nabla_r \Phi}|.
$
Here, $v_\phi$, $p$ and $\Phi$ are the rotation velocity, gas pressure 
and the gravitational potential, respectively.
The dashed lines show the ratio of the centrifugal force to the
radial gravitational force,
$
q_2=|\frac{v_\phi^2/x}{\nabla_r \Phi}|.
$
The dashed-dotted lines show $q=0.5$. In the regions where 
the dashed lines are
larger than the dashed-dotted lines, the gas is mainly 
supported by the centrifugal force.
The top and bottom panels show the results 
of Models Para and Ortho,respectively.
The epochs of each model are the same as those in figure \ref{xy_map}.
}
\label{ratio_rot}
\end{figure}

When the rotation vector and magnetic field are in 
the anti-parallel configuration, a large disk 
with a size of $r \sim 20 $ AU is formed 
(Model Para; top left panel).
The disk is so massive that spiral arms are created 
by the gravitational instability. 
We confirmed that the Toomre's $Q$ value, 
$Q=\kappa_{\rm ep} c_s/(\pi G \Sigma)$ is $Q\sim 1$ in the entire disk region
($5 \lesssim r \lesssim 20$ AU). 
In the top panel of figure \ref{ratio_rot}, we show the 
force balance between the pressure gradient force, 
the centrifugal force, and the radial gravitational force on the x-axis.
The gas is mainly supported by the centrifugal force.
Therefore, a rotationally-supported massive disk is formed
in Model Para.
On the other hand, when the rotation vector and magnetic 
field are in the parallel configuration, no large 
disk ($r \gtrsim 10$ AU) appears (Model Ortho; top right panel) 
because the magnetic braking is strengthened by the Hall 
current term and the angular momentum is 
efficiently removed from the central region. 
The dense region ($\rho>10^{-11} \cm$) at the 
center, which has a radius of $r\sim 5$ AU, is the remnant 
of the first core and is not a rotationally-supported disk. 
Although a rotationally-supported disk 
with $r \lesssim 0.6$ AU is also formed in Model Ortho 
around the protostar, 
as shown in the bottom panel of the figure \ref{ratio_rot};
the difference of the disk size of Model 
Para and Model Ortho is remarkable.


The plasma $\beta$ in the disk regions ($r \lesssim 20$ AU) 
of Model Para is large ($\beta>100$) because
the magnetic flux is largely removed due to the magnetic diffusion.
In this high $\beta$ region, the magnetic field and the 
gas are almost decoupled and
the magnetic braking is no longer important. 
The disk is sufficiently massive 
and develops gravitational instability. 
Thus, gravitational instability may play an important role 
for angular momentum transfer in the subsequent evolutionary phase.
Furthermore, in such a massive extended disk, disk fragmentation,
which is a promising mechanism for the formation of 
binaries or wide-orbit planets, 
\citep{2010Natur.468.1080M} possibly occurs in the subsequent evolution 
\citep[][]{1997Sci...276.1836B,2010ApJ...718L..58I,2011ApJ...729...42M,2011MNRAS.416..591T,2013MNRAS.428.1321T,2013MNRAS.436.1667T,2015MNRAS.446.1175T}. 
Thus, the parallel or anti-parallel property of the cloud core 
would play a crucial role for the formation 
of the binary or wide-orbit planets. 

\begin{figure}
\includegraphics[width=50mm,angle=-90]{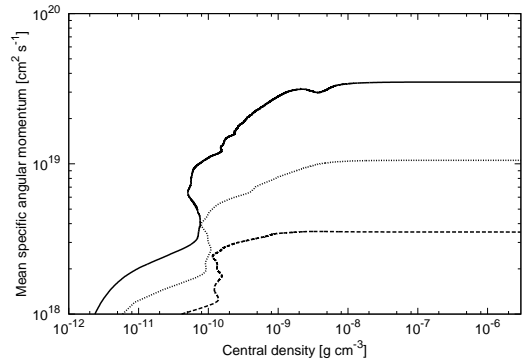}
\caption{
The time evolution of the mean specific angular momentum 
of the inner region with $\rho>10^{-12} \cm$ 
as a function of the central density. 
The solid, dashed, and dotted lines show 
the results of Model Para, Ortho, and NoHall, respectively.
}
\label{rho_J}
\end{figure}

To quantify the strength of the rotation at the center of the 
cloud core, we show the mean specific angular 
momentum of regions with $\rho>10^{\rm -12} \cm$ as a 
function of the central density in figure \ref{rho_J}. 
The figure shows that the specific angular momentum 
in Model Para is about an order of magnitude 
larger than that of Model Ortho. 
The specific angular momentum in Model Ortho (Para) is 
about three times smaller (larger) than that in Model NoHall. 
The combination of the spin-up effect 
(weakening of the magnetic braking) in the anti-parallel case 
and the spin-down effect (strengthening of the magnetic braking)
in the parallel case causes the large difference.

The mass and the magnetic flux 
$\Phi=\int {\mathbf B} d{\mathbf S}$ 
of regions with $\rho>10^{\rm -12} \cm$ in model 
Para, Ortho, and NoHall at the 
beginning of the second collapse ($\rho_c= 10^{-8} \cm$), 
was ($M ~(M_\odot),~\Phi ~{\rm (G~cm^2)}$)
=($1.9\times10^{-1},3.7\times 10^{28}$),
($7.5\times10^{-2}, 4.1\times 10^{27}$), 
and ($1.1\times10^{-1},8.0\times10^{27}$), respectively.
Here, $d{\mathbf S}$ is defined at the $z=0$ and is parallel to the z-axis.
Thus, the mass-to-flux ratio of the region 
normalized by its critical value $(M/\Phi)_{\rm crit}=(0.53/(3 \pi))(5/G)^{1/2}$ 
is $\mu=21,~74,$ and $~56$, respectively.
These values are much larger than the initial mass-to-flux ratio $\mu=4$.

\begin{figure}
\includegraphics[width=60mm,angle=-90]{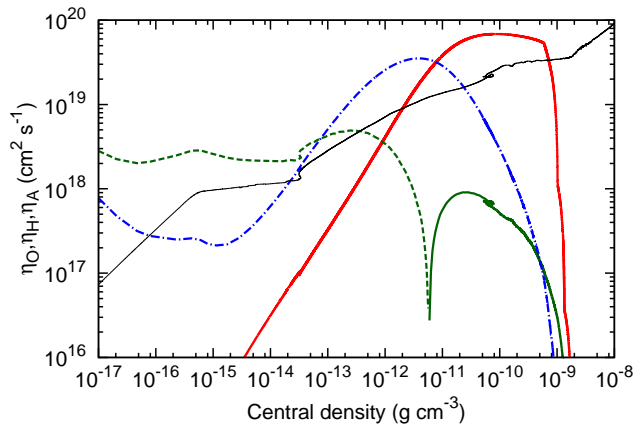}
\caption{
Magnetic diffusion coefficients , $\eta_O,~\eta_H,$ and $\eta_A$ at the center
as a function of the central density in Model Para.
The red line shows $\eta_O$,
The green line shows $\eta_H$ where the dashed line shows the region of $\eta_H<0$  and the solid line shows the region of $\eta_H>0$, and the 
blue dashed-dotted line shows $\eta_A$.
The black thin line shows the ``critical value'' 
of $\eta_H$, $3\times 10^{20}~B_c ({\rm cm^2 s^{-1}})$ suggested by
\citet{2011ApJ...733...54K} above which the disk is formed in their simulations.
Here, $B_c$ is the central magnetic field.
}
\label{eta}
\end{figure}

In figure \ref{eta}, we show the evolution of the 
Ohmic, Hall, and ambipolar diffusion coefficients,
 $\eta_O,~\eta_H,$, and $\eta_A$, at the center of model Para
as a function of the central density. 
The evolution of model Ortho and NoHall were almost the same.
In $\rho_c<10^{-14} \cm$, $\eta_H$ 
is larger than $\eta_A$ and $\eta_O$ and the
gas rotation is significantly affected by the Hall current term in this
region.
The value of $\eta_H$ is much larger than the ``critical 
value" for disk formation (thin black line) 
suggested by \citet{2011ApJ...733...54K} in this region.
Although the $\eta_H$ decreases in $\rho_c\gtrsim 10^{-13} \cm$,
the Ohmic and ambipolar diffusion alternatively 
play a role in $\rho_c\gtrsim 10^{-13} \cm$, and 
the rotation is maintained in the high density region 
without magnetic braking. 
Note that \citet{2011ApJ...733...54K} only considered the Hall term
and neglected other non-ideal effect.
We remark that 
$\eta_A$ does not strongly depend on $|\mathbf{B}|$ 
around $\rho \sim 10^{-14}~\cm$ 
although $\eta_A\propto |\mathbf{B}|^2$ in low density regions.

\begin{figure}
\includegraphics[width=90mm]{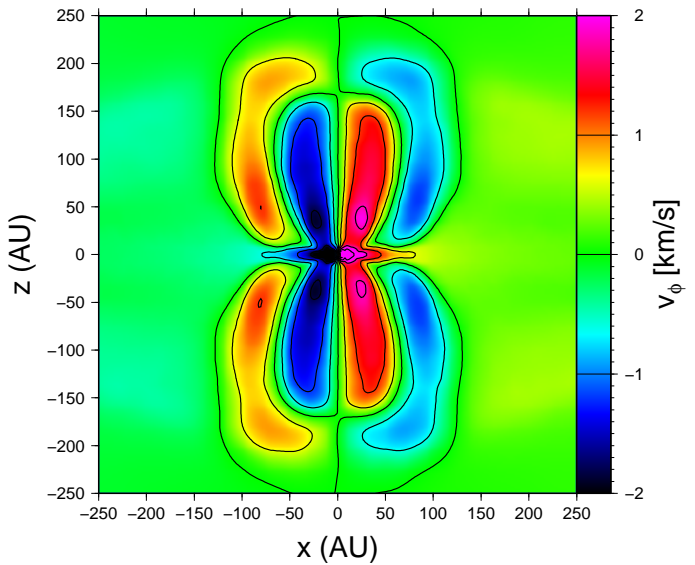}

\caption{
The cross section of $v_\phi$ in the 
x-z plane in Model Para.
The epoch of the snapshot is the same as that in Fig. \ref{xy_map}.
}
\label{vy_map}
\end{figure}

Because of the conservation of angular momentum, 
the spin-up due to the Hall term at the center causes spin-down 
of the outer region, eventually, causing anti-rotation against the disk.
In figure \ref{vy_map}, we show the cross 
section of the rotation velocity distribution of 
Model Para in the $x-z$ plane 
at the same epoch of figure \ref{xy_map}.
This figure clearly shows an anti-rotating 
envelope surrounding the forward rotating inner region.
Since the anti-rotation of envelope is driven 
by torsional Alfv\'{e}n waves, the anti-rotating region expands 
with time and will propagate to the outside 
of the parental core. 
Thus, the angular momentum of the direction 
opposite to the disk would 
eventually be cast away to the interstellar medium.


\section{Conclusions and Discussion}
In this study, we investigated the effect of 
the Hall current term on the formation of circumstellar disk.
by considering all non-ideal effects 
as well as the radiative transfer are considered.
To our knowledge, this is the first study that simultaneously 
includes these physical processes in a three-dimensional simulation.

We found that the disk evolution can 
be categorized into two cases depending whether
the magnetic field and the rotation vector are parallel or
anti-parallel. 
In the anti-parallel case, a relatively large ($r\gtrsim 10$ AU)
and massive disk forms simultaneously with the protostar formation;
however, a disk with $r\gtrsim 1$ AU does not form 
in parallel case.
Thus, the parity of the magnetic field significantly changes 
the disk formation process which has not been paid much attention to so far. 
Since the parallel and anti-parallel properties are 
inherent and do not readily change in time, we expect that 
the spin-up (the magnetic braking weakening) and spin-down 
(the magnetic braking strengthening) effects 
due to the Hall current term 
are also important in the subsequent disk evolution 
and the difference between the two cases is maintained or enhanced. 
Therefore, we suggest that the disk evolution 
can be categorized into two cases.
We may call the resultant parallel and anti-parallel systems as
{\it Ortho-disk} and {\it Para-disk}, respectively.

Our results predict that the bimodality in the disk size 
distribution spontaneously arises due to the Hall current 
term in typically magnetized molecular clouds.
We tend to think that the disk size has an unimodal 
distribution according to the 
strength of the rotation and the magnetic field of the cloud cores.
However, as we have shown above, the Hall term changes
the disk size according to the parallel or 
anti-parallel property of the cores. 
It is expected that about half of the molecular 
cloud cores have the parallel 
configuration and the others have the anti-parallel configuration
because the Hall term would
not play a role during the cloud core formation 
\citep{2004Ap&SS.292..317W} and there is no
physical mechanism which distinguishes the parallel 
and anti-parallel configurations.
On the other hand, during the gravitational collapse of cloud core, 
the Hall term becomes
effective and strengthens or weakens the magnetic braking.
Therefore, the bimodality of the disk size distribution 
spontaneously arises from 
the unimodal distributions of the rotation and the 
magnetic field strength of the cores.

An observational signature predicted
from our results is the anti-rotating envelope in the Class 0 YSOs.
At the end of the simulations, the anti-rotating envelope had a size of
$r\gtrsim 100$ AU and the rotation velocity of $v_\phi\sim 1$ km. 
Because the anti-rotation of the envelope stems from 
torsional Alfv\'{e}n waves, the anti-rotating region expands 
with time.
We predict future observations will find an anti-rotating envelope against 
disk rotation in the Class 0 YSOs. 
These observations will provide clear evidence  that 
the Hall current term plays an important role for the 
evolution of circumstellar disks.

In this paper, several simplifications were adopted and 
their influences should be investigated in future works.
We employed a fixed dust grain size of $a=0.035~ {\rm \mu m}$ and
a fixed cosmic-ray ionization rate of $\xi_{\rm CR}=10^{-17} s^{-1}$ and
the magnetic resistivities are sensitive to the models of dust and cosmic-ray
\citep{2012A&A...541A..35D,2014A&A...571A..33P}. 
Furthermore, the drift velocity of the magnetic field induced by the 
Hall term which characterizes the strength of the Hall term,
linearly depends on $\eta_H$.
Thus, the simulations with the different models are 
necessary to confirm our results.
The misalignment between the magnetic field and the rotation vector is 
another important issue. In our simulations, 
the initial rotation vector and the magnetic
field are in perfectly parallel or anti-parallel configurations.
However, it is expected that they are mutually 
misaligned in the realistic cloud cores 
\citep{2009A&A...506L..29H,2012A&A...543A.128J}.
Its effect on the disk formation with non-ideal effects
should also be investigated.
We used a rigidly rotating shell as the outer boundary
condition.
Because of the angular momentum out-flux at the boundary, 
the total angular momentum is non-conserved quantity in our simulations.
At the end of the simulations,
the total angular momentum of Models Para, Ortho, and
NoHall within the boundary shell 
was $97,~94.3$, and $95.5$ \% of the initial angular momentum, respectively. 
Note that the difference of the angular momentum between models
is expected because the Hall current term 
changes the angular momentum transfer rate near the boundary.
Similar phenomena are also observed in the previous works.
Previous simulations starting from the non-rotating core 
with the Hall term and outgoing boundary have a finite angular 
momentum at the end of the 
simulations \citep[][]{2011ApJ...733...54K,2011ApJ...738..180L}.
We expect the treatment of the 
outer boundary condition would not change our results significantly
because the crossing time of Alfv\'{e}n wave $t_{\rm cross}=R/v_A$ 
is larger than the free-fall time $t_{\rm ff}$
(in our initial condition, $t_{\rm cross}/t_{\rm ff}=2.5$)
and the boundary mainly influences the relatively outer region 
within our simulation time $t_{\rm sim}\lesssim 1.1 t_{\rm ff}$.
However, more sophisticated boundary conditions are desired.

\section *{Acknowledgments}
We thank K. Tomida and Y. Hori for providing EOS table.
The computations were performed on the
XC30 system at CfCA of NAOJ.

\bibliography{article}

\end{document}